\documentclass[%
 reprint,
 amsmath,amssymb,
 aps,
showkeys
]{revtex4-2}

\usepackage{graphicx}
\usepackage{dcolumn}
\usepackage{bm}

\usepackage{xcolor}
\usepackage[separate-uncertainty=true]{siunitx}
\DeclareSIUnit{\rpm}{rpm}
\usepackage{hyperref}
\newcommand{\aref}[1]{\hyperref[#1]{Appendix~\ref*{#1}}}

\usepackage{todonotes}

\newcommand{\wttop}[1]{{#1}_\mathrm{top}}

\newcommand{\deltau}{\Delta \overline{u}_D / \overline{u}_\mathrm{top}}

\newcommand{\caseshearlow}{low shear}
\newcommand{\casetilow}{low turbulence intensity}
\newcommand{\casetsrlow}{low tip speed ratio}
\newcommand{\casebase}{reference case}
\newcommand{\casetsrhigh}{high tip speed ratio}
\newcommand{\casetihigh}{high turbulence intensity}
\newcommand{\caseshearhigh}{high shear}

\newcolumntype{P}{D{p}{\, \pm \,}{6,6}}

\begin{document}

\preprint{APS/123-QED}

\title{Effect of inflow conditions on tip vortex breakdown\\in a high Reynolds number wind turbine wake}

\author{Mano Grunwald}%
\affiliation{%
 Max Planck Institute for Dynamics and Self-Organization, 37077 Göttingen, Germany}%
 \affiliation{%
 Department of Physics, Georg-August-Universität Göttingen, 37077 Göttingen, Germany}%
 \author{Claudia E. Brunner}%
 \email{claudia.brunner@ds.mpg.de}%
 \affiliation{%
 Max Planck Institute for Dynamics and Self-Organization, 37077 Göttingen, Germany}%

\date{\today}

\begin{abstract}
Understanding the re-energization of wind turbine wakes is crucial for the design and control of wind farms. Close to the rotor, this process is determined by the dynamics of the tip vortices. Here, we experimentally investigate the downstream evolution of the tip vortices for different inflow conditions. The experiments were performed in the Variable Density Turbulence Tunnel at the Max Planck Institute for Dynamics and Self-Organization, which uses pressurized $\mathrm{SF}_6$ as the working fluid to achieve a turbine diameter-based Reynolds number of~$\mathrm{Re}_D=2.9\times10^6$. An active turbulence grid was used to generate atmospheric inflow conditions with varying levels of mean shear and turbulence intensity. Hot wire measurements of the streamwise velocity component were conducted in the inflow and the wake of a model wind turbine MoWiTO~0.6 for various tip speed ratios and are used to investigate the scaling of tip vortex breakdown in the near wake. While the scaling is only weakly affected by variations in mean velocity shear, both turbulence intensity and tip speed ratio have a strong effect on
vortex breakdown.
\end{abstract}

\maketitle

\section{\label{sec:introduction}Introduction}
Wind energy is a key technology in the transition to green energy production. Wind turbines are commonly organized in wind farms, where multiple turbines operate in close proximity. This leads to interactions between the turbines, with downstream turbines experiencing the wakes of upstream turbines as their inflow. These interactions reduce the generated power~\cite{barthelmie_quantifying_2010,hansen_impact_2012} and increase fatigue loads on the blades of the turbines~\cite{thomsen_fatigue_1999}. Therefore, wakes play a critical role in the development of wind farm layouts, necessitating a comprehensive understanding of their dynamics.

Wind turbine wakes are dominated by coherent structures: vortices are shed from roots and tips of the rotor blades and transported downstream. Due to the circular motion of the rotor, helical vortex tubes are formed. While the root vortices break down close to the rotor, the tip vortices can persist several rotor diameters downstream. It is their evolution that primarily determines the initial dynamics of the wake~\cite{vermeer_wind_2003,porte-agel_wind-turbine_2020,neunaber_distinct_2020}. In the near wake, the velocity is low as the wind turbine removes kinetic energy from the flow. Only once the tip vortices break down, does the wake begin to re-energize~\cite{lignarolo_entrainment_2014,lignarolo_experimental_2014}.
Mechanisms of tip vortex breakdown have been studied extensively in the literature and can be grouped into two categories: interactions with adjacent vortices and interactions with surrounding turbulence. Which mechanisms dominate is likely dependent on the exact flow conditions. As such, understanding what mechanisms induce vortex breakdown under realistic atmospheric conditions and what scaling laws they follow is of great interest to the wind energy community.

Breakdown due to interactions with adjacent vortices is the most widely studied mechanism.
Ivanell~et~al.~\cite{ivanell_stability_2010} conducted Large Eddy Simulations to investigate the stability of the tip vortices by perturbing them in the axial direction. They found that tip vortex breakdown is faster if subsequent helices are perturbed in different directions. This brings the vortices closer together, which promotes interactions between them.
While the interaction of vortex pairs has been extensively studied at a fundamental level (see e.g.~\cite{leweke_dynamics_2016}), Posa~et~al.~\cite{posa_instability_2021} later described the breakdown mechanism of tip vortices in wind turbine wakes due to interaction in detail. After short-wave instabilities are triggered, long-wave instabilities lead to mutual induction between subsequent vortices. This eventually results in leapfrogging and the breakdown of the tip vortices into smaller eddies.

The spacing of adjacent vortices is related to the tip speed ratio~TSR:
\begin{equation}
    \mathrm{TSR}=\frac{\pi D f_\mathrm{rot}}{\overline{u}}
\end{equation}
Here, $D$ is the rotor diameter, $f_\mathrm{rot}$ is the rotation frequency of the rotor and $\overline{u}$ is the mean velocity of the flow. The rotation rate of the rotor determines how often vortices are created at a given location, while the mean velocity determines how quickly they are advected downstream. As such, higher tip speed ratios result in closer vortices. 
%
Hu~et~al.~\cite{hu_dynamic_2012} point to a decrease in tip vortex strength with increasing tip speed ratio due to the decreasing effective angle of attack of the blades. As a result, they report faster vortex breakdown at higher tip speed ratios. Thus, higher tip speed ratios not only enhance the breakdown process by decreasing the distance between vortices, but also by changing the effective angle of attack. They also record the maximum vorticity of the tip vortices as a function of downstream distance from the rotor. Interestingly, they find that the decrease in vorticity follows a power law.

In the presence of turbulence, the interaction of tip vortices with the surrounding flow is also thought to induce vortex breakdown.
%
Ghimire and Bailey~\cite{ghimire_experimental_2017} found that higher turbulence intensity in the inflow leads to a faster breakdown of the tip vortices. In a subsequent study, they showed that a turbulent inflow led to fluctuations in the vortex core circulation, which were absent under laminar inflow conditions~\cite{ghimire_experimental_2018}. They found that these fluctuations correspond to the formation of secondary vortex structures around the tip vortices. Interactions with these structures then led to breakdown of the tip vortices.

In the context of wind turbines, the turbulence intensity (TI) is typically used to characterize the inflow turbulence:
\begin{equation}
    \mathrm{TI}=\frac{u'_\mathrm{rms}}{\overline{u}}
\end{equation}
Here, $u'_\mathrm{rms}$ is the root-mean-square value of the velocity fluctuations in the streamwise direction~$u'$.
%
Gambuzza and Ganapathisubramani~\cite{gambuzza_influence_2023} showed that higher turbulence intensities lead to faster wake recovery.
Van der Deijl~et~al.~\cite{van_der_deijl_effect_2024} found that this effect is stronger than the effect of tip speed ratio.
%
Yen~et~al.~\cite{yen_near_2024} explicitly studied the leapfrogging effect, by shortening one of the rotor blades. Although this did lead to enhanced interactions between the tip vortices at low TI, interactions with the surrounding turbulence dominated vortex breakdown at higher TI.

The inflow experienced by real wind turbines is typically not only turbulent, but also sheared. However, few studies have investigated the impact of shear on tip vortex breakdown.
%
%
Parinam~et al.~\cite{parinam_large-eddy_2023} conducted a Large Eddy Simulation of a wake in a sheared laminar inflow and observed a tilting of the helical tip vortex structures due to the shear. This led to differences in the spacing of adjacent vortices: at the bottom of the rotor, tip vortices were closer together and thus interacted sooner, leading to faster breakdown, while at the top of the rotor, they were spaced further apart, delaying interactions and breakdown.

Because vortex breakdown is sensitive to the local flow conditions, it is crucial to investigate these effects under conditions that are as realistic as possible. However, the Reynolds numbers of wind turbine flows are exceptionally high and therefore challenging to recreate in numerical studies and laboratory experiments. The turbine-based Reynolds number~$\mathrm{Re}_D=\overline{u}D/\nu$ is typically on the order of $10^6\lesssim \mathrm{Re_D} \lesssim 10^8$. However, most laboratory experiments of wind turbine wakes have been conducted at Reynolds numbers that are more than two orders of magnitude lower than those of full-scale wind turbines.
%
An exception to this is the use of pressurized wind tunnels, which generate high Reynolds number flows by lowering the kinematic viscosity. Piqué~et~al.~\cite{pique_laboratory_2022} conducted hot-wire measurements in such a facility to study the wake of a model wind turbine and observed a Reynolds number invariance of the wake velocity field over $2.7\times10^6\leq \mathrm{Re}_D \leq 7.2\times 10^6$. In a subsequent study, they investigated coherent structures in the wake by associating them with the presence of distinct peaks in the velocity spectra~\cite{pique_dominant_2022}. The peak corresponding to the tip vortices decreased with increasing downstream position. Importantly, these experiments were conducted under laminar inflow conditions, so that the effect of inflow turbulence could not be studied.

Here, we present an experimental investigation of tip vortex breakdown in turbulent, sheared flows at full-scale Reynolds numbers. These experiments were conducted in the Variable Density Turbulence Tunnel (VDTT) at the Max Planck Institute for Dynamics and Self-Organization to achieve a Reynolds number of~$\mathrm{Re}_D=2.9\times 10^6$. An active turbulence grid was used to control mean velocity shear and turbulence intensity in the inflow. To the authors' knowledge, this is the first laboratory study of a wind turbine in a high Reynolds number~($\mathrm{Re}_D\sim 10^6$) turbulent flow. The focus of the present study is not to identify which specific mechanisms initiate vortex breakdown, but rather the scaling of the breakdown process in turbulent shear flows. As such, we present the effects of turbulence intensity, mean shear and tip speed ratio on the signatures of the tip vortices in the velocity spectra.  

\begin{figure}
    \centering
    \includegraphics[width=0.95\columnwidth]{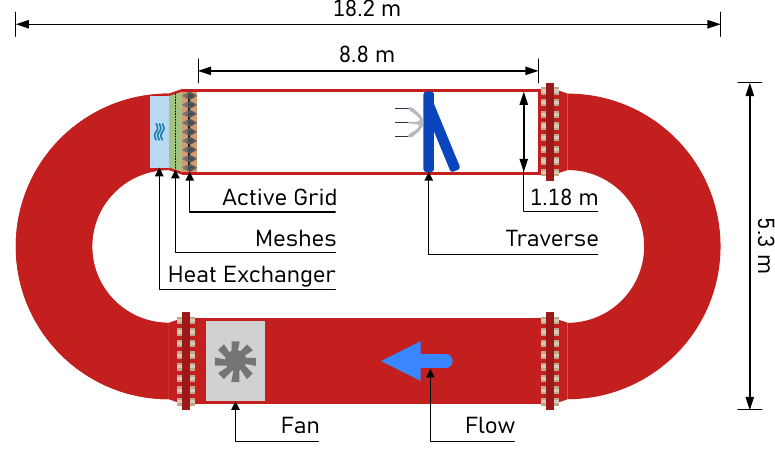}
    \caption{Schematic of the Variable Density Turbulence Tunnel at the Max Planck Institute for Dynamics and Self-Organization.}
    \label{fig:vdtt}
\end{figure}
\section{\label{sec:methodology}Methodology}
\subsection{\label{sec:methodology:vdtt}Variable Density Turbulence Tunnel}
The experiment was conducted in the Variable Density Turbulence Tunnel (VDTT) at the Max Planck Institute for Dynamics and Self-Organization in Göttingen. The VDTT, shown in~\autoref{fig:vdtt}, is a closed-loop wind tunnel in an upright position.
It is~\SI{18.2}{\metre} long,~\SI{5.3}{\metre} tall and contains an upper and a lower measurement section. The present experiments were conducted in the~\SI{8.8}{\metre} long upper measurement section. The VDTT is specifically designed to investigate high-Reynolds-number flows. To this end, sulfur hexafluoride ($\mathrm{SF}_6$) at up to~\SI{15}{\bar} is used as the working fluid due to its low kinematic viscosity. A~\SI{210}{\kilo\watt} fan located at the downstream end of the lower measurement section is used to circulate the gas at up to~\SI[per-mode=symbol]{5}{\metre\per\second}. The experimental conditions of the present study are shown in~\autoref{tab:experimental_conditions}.
\begin{table}
    \centering
    \begin{ruledtabular}
        \begin{tabular}{lll}
          variable name & symbol & mean value \\
         \hline\\
         mean velocity & $\wttop{\overline{u}}$ & $\SIrange[]{2.87}{2.94}{\metre\per\second}$\\
         pressure & $p$ & \SI{4.00}{\bar}\\
         temperature & $T$ & \SI{20.0}{\celsius}\\
         density & $\rho$ & \SI{25.2}{\kilogram\per\metre\cubed}\\
         dynamic viscosity & $\mu$ & \SI{15.0}{\micro\pascal\second}
        \end{tabular}
    \end{ruledtabular}
    \caption{Experimental conditions. The mean velocity is obtained at the upper edge of the rotor.}
    \label{tab:experimental_conditions}
\end{table}
Before the working fluid enters the upper measurement section, it passes through a heat exchanger to ensure constant temperature throughout the experiment. The heat exchanger additionally acts as a flow straightener that removes any fan rotation from the flow. Downstream of the heat exchanger three meshes with increasing mesh sizes are installed to laminarize the flow. The wind tunnel is described in detail in Bodenschatz et al.~\cite{bodenschatz_variable_2014}.

\begin{figure}
    \centering
    \includegraphics[width=0.9\columnwidth]{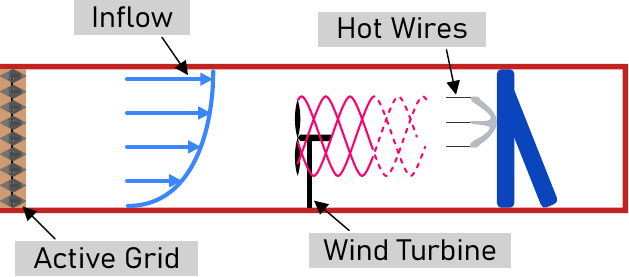}
    \caption{Experimental setup in the upper measurement section of the VDTT.}
    \label{fig:setup}
\end{figure}
The experimental setup is shown in~\autoref{fig:setup}. An active turbulence grid is located immediately downstream of the laminarizing meshes. A customary inflow profile is generated with the active grid and the streamwise velocity component of the flow is recorded at a variety of locations using a hot wire setup. After characterizing the inflow, a model wind turbine is installed and the streamwise velocity component at the top of the wake is measured at various downstream positions. The following sections provide detailed information on each of these steps.

\subsection{\label{sec:methodology:generating_inflow_conditions}Active turbulence grid}
To investigate the impact of shear, turbulence intensity and tip speed ratio on tip vortex breakdown, several inflow cases are generated with the active grid. The use of active grids for generating shear flows~\cite{cekli_tailoring_2010, hearst_tailoring_2017, jooss_influence_2023} and turbulence~\cite{makita_realization_1991, hearst_tailoring_2017, knebel_atmospheric_2011, neuhaus_generation_2020, griffin_control_2019, kroger_generation_2018} is well established in the literature. The active grid in the VDTT consists of~111 flaps, which can be individually controlled at a rate of~\SI{10}{\hertz}. Each flap is quadratic with a side length of~\SI{10}{\centi\metre}. The gap between two fully closed consecutive flaps is~\SI{15}{\milli\metre}, resulting in a mesh size of $M = \SI{11.5}{\centi\metre}$. The flaps are arranged in 9 rows and 13 columns. A flap at row~$i$ and column~$j$ is characterized by its opening angle~$\phi_{ij}$ with~$\phi_{ij}=0$ corresponding to a fully open flap and $\phi_{ij}=\pm\SI{90}{\degree}$ corresponding to a fully closed flap. The grid motion is composed of a stationary mean distribution~$\overline{\phi}_{ij}$ that is superimposed with a random fluctuating motion~$\phi'_{ij}(t)$ that is a function of time~$t$, such that $\phi_{ij}(t) = \overline{\phi}_{ij} + \phi'_{ij}(t)$. The time series for the random motion is integrated from an Ornstein-Uhlenbeck process following Neuhaus et al.~\cite{neuhaus_generation_2020}:
\begin{equation}
    \frac{\mathrm{d}\phi'_{ij}}{\mathrm{d}t}(t) = -\gamma_\mathrm{ou}\phi'_{ij}(t)+\sqrt{D_\mathrm{ou}}\,\Gamma(t)
\end{equation}
Here, $\gamma_\mathrm{ou}$ determines how strongly the process is attracted to zero, $D_\mathrm{ou}$ governs the magnitude of the random fluctuations and $\Gamma$ is noise drawn from a Gaussian distribution. Additionally, spatial correlation is introduced following the principles described in Griffin et al.~\cite{griffin_control_2019}:
\begin{equation}
    \phi'_{ij} = \sum\limits_{k=1}^{3}\sum\limits_{l=1}^{3}\Omega_{kl}\phi'_{i+k-2,j+l-2}
\end{equation}
Here, a 3x3 convolution with a convolution kernel~$\Omega_{kl} = 1/9$ is used. At the boundaries, reflective boundary conditions are employed such that $\phi'_{(-1)j}=\phi'_{1j}$ and $\phi'_{i(-1)}=\phi'_{i1}$. Using this kernel, the angle of each flap is determined by the average of its own angle and the angles of its eight neighboring flaps.

Target velocity profiles with varying levels of shear and turbulence intensity were determined. The parameters of the above grid motion protocol were then iteratively adjusted until satisfactory agreement between the resulting flow and the target profiles was achieved. As the target mean velocity shear is in the vertical direction, the mean flap angles are considered to be the same across each row such that~$\overline{\phi}_{ij} = \overline{\phi}_i$. The optimization parameters were the mean angle of each row $\overline{\phi}_i$, $D_\mathrm{ou}$ and the wind tunnel fan speed. The parameters used for the final grid protocols are presented in~\aref{sec:appendix_grid}. More details on the active grid are provided in Bodenschatz et al.~\cite{bodenschatz_variable_2014} and Griffin et al.~\cite{griffin_control_2019}.

\subsection{\label{sec:methodology:mowito}Model wind turbine}
The wind turbine used in these experiments is a model wind turbine MoWiTO~0.6, developed at the University of Oldenburg~\cite{schottler_design_2016, juchter_reduction_2022}. The turbine is placed~$40.7M$ downstream of the active grid. It has a rotor diameter of~$D=\SI{0.58}{\metre}$ and a hub height of~$z_\mathrm{hub}=\SI{0.52}{\metre}$. The conditions shown in Fig.~\autoref{tab:experimental_conditions} result in a turbine Reynolds number of~$\mathrm{Re_D} = \SI{2.9e6}{}$. The blade design is based on the SD~7003 airfoil. The turbine features a pitching mechanism, allowing for the simultaneous adjustment of the pitch of the rotor blades. The rotor is connected to a DC~motor that acts as a generator. The turbine rotation rate is controlled by a field effect transistor (FET) which dissipates the generated power. The voltage of the FET is adjusted using a PID controller. The increased power density of the flow in $\mathrm{SF}_6$ compared to air means that at a given rotation rate the rotor extracts more power. At high pressures, this power exceeded the maximum power the generator could convert at this rotation rate. As a result, it was not possible to apply the load necessary to fix the rotation rate and the rotor would spin up. A gearbox with a ratio of $1:4.8$ was installed to reduce the rotation rate of the generator and thus increase its torque. This shifted the operating conditions closer to the optimal range for this generator, so that it was able to extract the necessary power.

\subsection{\label{sec:methodology:data_aquisition}Data acquisition and post-processing}
All velocity information in this study is based on streamwise velocity measurements conducted using Dantec Dynamics 55P11 hot wire probes. The probes are~\SI{1.25}{\milli\metre} long with a diameter of~\SI{5}{\micro\metre}. The hot wires were operated in CTA-mode and controlled via a Dantec Dynamics StreamLine~90N10 Frame with 90C10 CTA modules using the Dantec Dynamics StreamWare~Pro~v6.00 software. The overheat ratio was set to \SI{0.8}{}. For the connection to the modules, \SI{20}{\metre} long RG223 cables were used. The signal was sampled at \SI{60.06}{\kilo\hertz}, low pass filtered at \SI{30}{\kilo\hertz} and read out via a National Instruments PCI-6123 DAQ card. The hot wires were calibrated using pitot tubes connected to a Siemens SITRANS~P~DS~III differential pressure transducer. Turbine data, such as the rotation rate, were sampled at~\SI{5}{\kilo\hertz} by a National Instruments cRIO-9074 data acquisition system.

The inflow profiles are characterized at the position of the model wind turbine in the lateral center of the wind tunnel with a vertical spacing of~\SI{24}{\milli\metre} between measurement points. The measurements extend beyond the swept area of the rotor as shown in~\autoref{fig:inflow_profiles}. At each measurement point, the flow is sampled for~\SI{5}{\minute} to ensure convergence of second moments. To verify convergence, \SI{15}{\minute} measurements are conducted in the lateral and vertical center of the wind tunnel for each grid protocol.

\begin{figure}
    \includegraphics[width=\columnwidth]{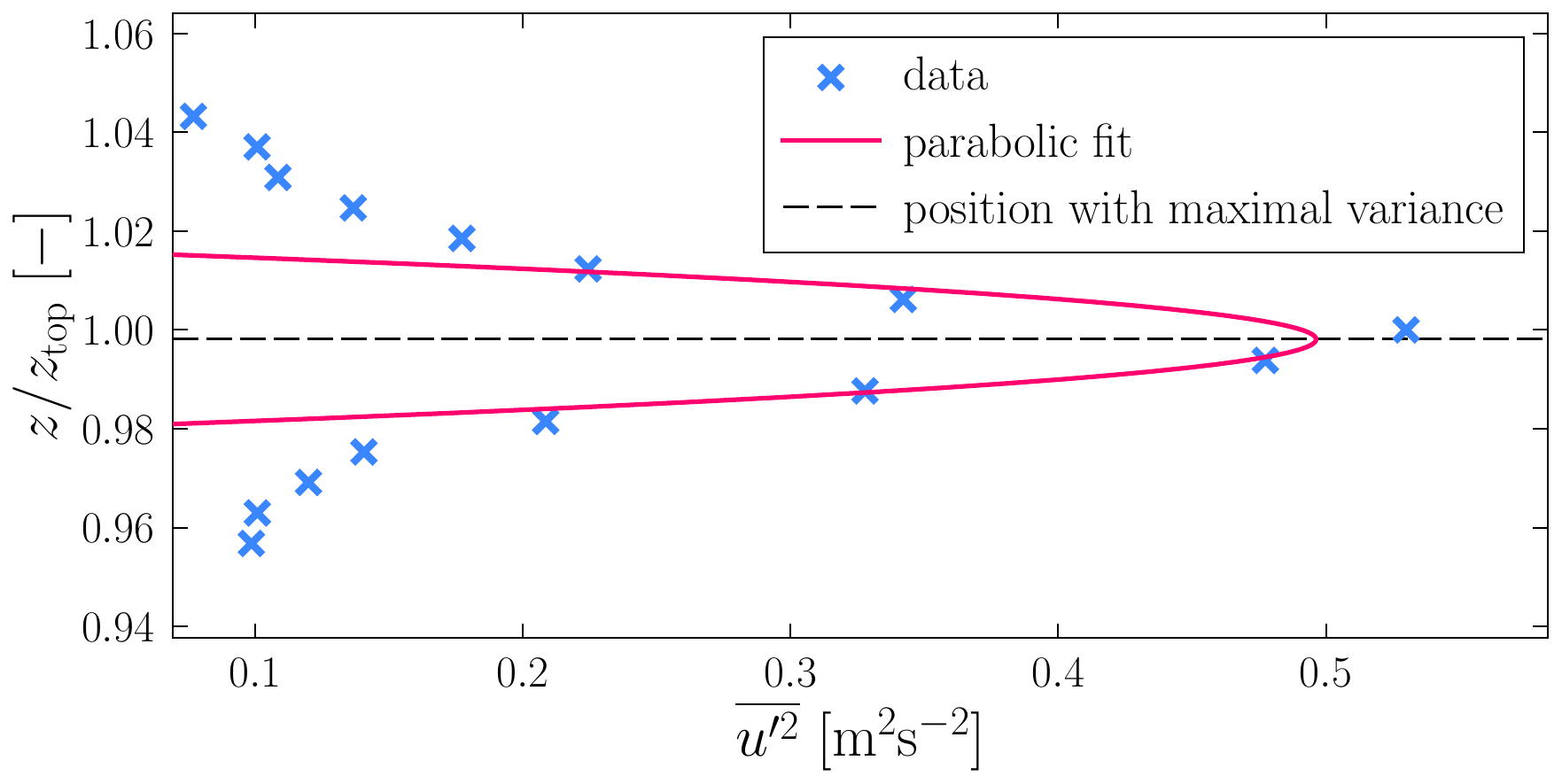}
    \caption{\label{fig:variance}Streamwise velocity variance profile at the top of the wake at $x/D=0.25$ downstream of the rotor. The maximum of the parabolic fit is taken to be the tip vortex height, at which a longer measurement is then conducted to determine the streamwise velocity spectrum.}
\end{figure}

To observe the tip vortex evolution in the wake, measurements were conducted at the lateral center of the tunnel at streamwise positions~$x/D = [0.25, 0.50, 0.75, 1.00, 1.25, 1.50, 6.79]$ downstream of the model wind turbine. Due to wake expansion, the vertical location of the tip vortices is a function of downstream position~\cite{jensen_note_1983, bastankhah_new_2014}. To account for this, the vertical profile of the velocity variance was measured throughout in the expected region following~\cite{pique_laboratory_2022}. An example of such a variance profile is shown in~\autoref{fig:variance}. A parabola is fit to this profile and its maximum is taken to be the tip vortex height. The flow was sampled at this position for~\SI{15}{\minute} resulting in the streamwise velocity spectra presented in~\autoref{sec:results}. In the two most downstream positions, the tip vortices had typically broken down so much that the variance profiles no longer showed clear maxima. In these cases, the height of the closest upstream measurement point was used.

The streamwise velocity spectra are computed using Welch's algorithm~\cite{welch_use_1967} and smoothed with a moving average filter. This procedure is parameterized by the number of segments used in Welch's algorithm and the width of the moving average window. Here, eight segments are used, while the width of the moving average window is~\SI{0.3}{\hertz}. Details on these parameter choices can be found in~\aref{sec:appendix_parameters}. Typically, turbulent velocity spectra follow a $-5/3$ scaling in the inertial range. In the inflow, this scaling extends over three decades which is indicative of a high Reynolds number flow.

\subsection{\label{sec:methodology:inflow_characterization}Inflow characteristics}
In a vertically-sheared flow, many relevant parameters inherently vary as a function of height. This required us to select a reference height at which to define target inflow characteristics. We decided to focus on the tip vortices shed from the upper edge of the rotor because these are easiest to locate and least affected by the turbine tower. As such, the upper edge of the rotor was chosen as the reference location for defining the inflow conditions, rather than the hub height. To minimize $\mathrm{Re}_D$ effects on the tip vortices, the target mean velocity at the top of the rotor is the same for all cases. Due to the presence of boundary layers at the tunnel walls, maintaining a linear mean velocity gradient across the entire vertical profile is not possible. However, we expect the behavior of the tip vortices to be primarily sensitive to the velocity gradient across the rotor area. As such, the mean shear is here defined as the mean velocity difference~$\Delta \overline{u}_D$ across the vertical span of the rotor normalized by the mean velocity at the upper edge of the rotor:
 \begin{equation}
     \frac{\Delta \overline{u}_D}{\overline{u}_\mathrm{top}} = \frac{D}{\overline{u}_\mathrm{top}m}
 \end{equation}
Here, $m$ is the slope of a linear fit to the recorded mean velocity profiles over the vertical span of the rotor and $\overline{u}_\mathrm{top}$ is the mean velocity at the upper edge of the rotor. Because turbulence in the inflow is generated not only by the active grid, but also by the mean velocity shear, it is practically impossible to keep the turbulence intensity constant across the vertical profile of the test section. However, we expect the behavior of the tip vortices at the upper edge of the rotor to be primarily sensitive to the turbulence intensity at their height. As such, the turbulence intensity is referenced at the height of the upper edge of the rotor $\mathrm{TI}_\mathrm{top} = (u'_\mathrm{rms})_\mathrm{top}/\overline{u}_\mathrm{top}$. Equivalently, the tip speed ratio is defined based on the velocity at the top of the rotor $\mathrm{TSR}_\mathrm{top} = \pi D f_\mathrm{rot}/\overline{u}_\mathrm{top}$.

\begin{figure*}
    \includegraphics[width=\textwidth]{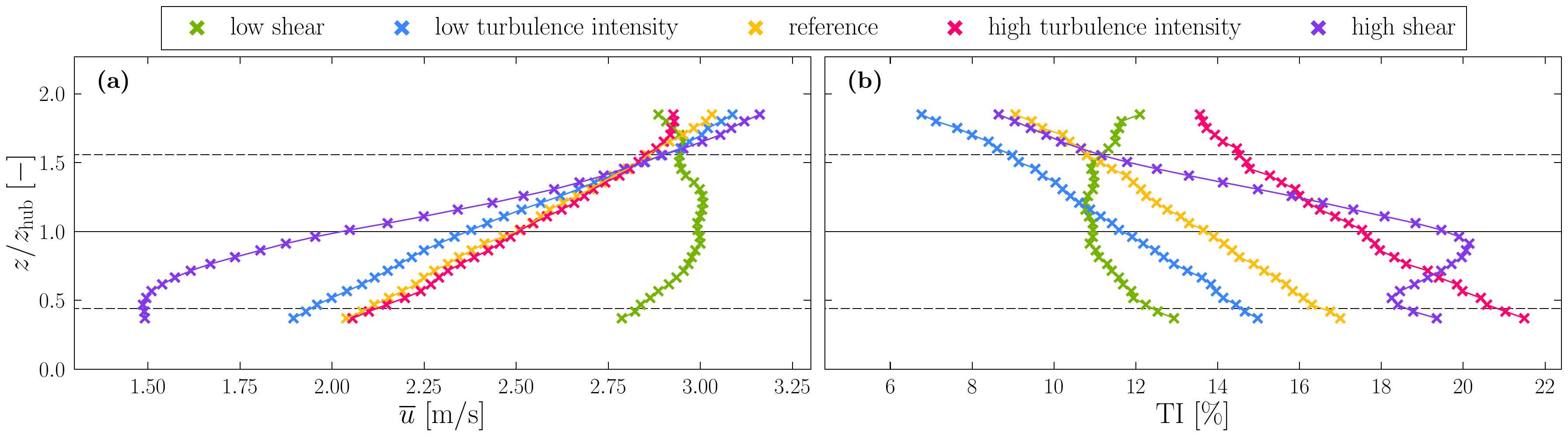}
    \caption{\label{fig:inflow_profiles}(a) mean velocity and (b) turbulence intensity profiles. Data with the same color in both plots corresponds to the same inflow case. The solid black line marks the hub height of the model wind turbine, while the dashed black lines show the rotor swept area.}
\end{figure*}

The resulting mean velocity and turbulence intensity profiles for the five investigated inflow conditions are shown in~\autoref{fig:inflow_profiles} and the relevant parameters are listed in~\autoref{tab:inflow}. 
\begin{table*}
    \begin{ruledtabular}
        \begin{tabular}{ldddddc}
            ID & \multicolumn{1}{c}{$\deltau$} & \multicolumn{1}{c}{$\mathrm{TI}_\mathrm{top}$ $[\si{\percent}]$} & \multicolumn{1}{c}{$\mathrm{TSR}_\mathrm{top}$} & \multicolumn{1}{c}{$L_{11}$ $[\si{\metre}]$} & \multicolumn{1}{c}{$\eta$ $[\si{\micro \metre}]$} & \multicolumn{1}{c}{$\mathrm{Re}_\lambda$} \\
            \hline
            \caseshearlow   & 0.10 & 11.3 & 6.1 & 0.67 & 41 & 1850  \\
            \casetilow      & 0.33 &  8.8 & 6.2 & 1.01 & 51 & 1890  \\
            \casetsrlow     & 0.27 & 10.6 & 4.2 & 0.81 & 50 & 2510  \\
            \casebase       & 0.27 & 10.6 & 6.3 & 0.81 & 50 & 2510  \\
            \casetsrhigh    & 0.27 & 10.6 & 8.4 & 0.81 & 50 & 2510  \\
            \casetihigh     & 0.26 & 14.6 & 6.3 & 0.60 & 35 & 2570  \\
            \caseshearhigh  & 0.57 & 10.8 & 6.2 & 0.64 & 44 & 2030
        \end{tabular}
    \end{ruledtabular}
    \caption{Mean shear~$\deltau$, turbulence intensity~$\mathrm{TI}_\mathrm{top}$, tip speed ratio~$\mathrm{TSR}_\mathrm{top}$, integral length scale~$L_{11}$, Kolmogorov scale~$\eta$ and Taylor scale Reynolds number~$\mathrm{Re}_\lambda$ for all generated inflow cases. All values are measured at the height of the upper edge of the rotor.}
    \label{tab:inflow}
\end{table*}
The profiles are smoothed with a moving average filter. The case of medium shear and medium turbulence intensity serves as a reference case. Two cases differ in their turbulence intensity but have approximately the same mean shear across the rotor plane, while the two remaining cases have varying mean shear but approximately the same turbulence intensity at the top edge of the rotor. As such, these five inflow conditions allow us to study the effects of shear and turbulence on tip vortex breakdown independently of one another. The effect of tip speed ratio was investigated using the reference case inflow condition and varying the tip speed ratio over $\mathrm{TSR}_\mathrm{top} = [4.2, 6.3, 8.4]$. The parameter space studied here falls within the realistic operating conditions of wind turbines with the turbulence intensity being on the higher end of the range.

To further characterize the inflow turbulence of the different cases, the integral length scale, the Kolmogorov length scale and the Taylor-scale Reynolds number are shown. For the computation of the turbulence scales, Taylors frozen flow hypothesis is employed to convert temporal scales into length scales. It should be noted that this hypothesis is only valid for small turbulence intensities and may introduce some error in the derived quantities for the highly turbulent flows investigated here. For all following relations see~\cite{pope_turbulent_2000}. The integral length scale~$L_{11}$ is obtained by numerically integrating the autocorrelation function of the streamwise velocity component until the first zero crossing. The Kolmogorov length scale~$\eta$ is defined as $\eta=(\nu^3/\varepsilon)^{1/4}$. The dissipation rate~$\varepsilon$ is obtained via the streamwise velocity spectrum~$E_{11}$. By compensating the spectrum with the Kolmogorov scaling~$E_{11}\sim \varepsilon^{2/3}f^{-5/3}/\overline{u}$, where $f$ is a frequency, the dissipation rate is computed as an average over the inertial subrange~\cite{schroder_estimating_2024}:
\begin{equation}
    \varepsilon = \left\langle\frac{2\pi}{\overline{u}}\left[\frac{f^{5/3} E_{11}}{18/55C_K}\right]^{3/2}\right\rangle_{\text{inertial range}}
\end{equation}
However, this method is based on the assumption of homogeneous isotropic turbulence which is likely not valid for the flow conditions involving mean velocity shear. The Taylor scale Reynolds number~$\mathrm{Re}_\lambda$ is defined as $\mathrm{Re}_\lambda = u'_\mathrm{rms}\lambda/\nu$, where $\lambda$ is the Taylor microscale. This Reynolds number is commonly used to characterize grid turbulence. The Taylor scale is obtained by
\begin{align}
    \lambda = \sqrt{\frac{15\nu {u'}_\mathrm{rms}^2}{\varepsilon}}
\end{align}

The Reynolds number of grid turbulence is generally considered to be high if ~$\mathrm{Re}_\lambda\sim10^3$, which is achieved for all cases. It is somewhat surprising that the high-turbulence-intensity inflow had approximately the same $Re_\lambda$ as the reference case and that the high-shear inflow had a lower $Re_\lambda$ than the reference case. 
In all cases, the integral scale is equal to or larger than the rotor diameter, which is generally also true for full-scale wind turbine flows. While a dependence of the integral scale on mean shear is not observed, it decreases with increasing turbulence intensity. 
Similarly, the Kolmogorov scale $\eta$ decreases with turbulence intensity.

\subsection{\label{sec:methodology:quantifying_tip_vortex_strength}Quantification of tip vortex breakdown}
In order to investigate the scaling of tip vortex breakdown, a quantitative measure of this breakdown is needed. Many studies use vorticity-based measures derived from flow visualization to quantify vortex strength~\cite{hu_dynamic_2012, lignarolo_experimental_2014}. However, such approaches are time-consuming and computationally expensive, which makes large parameter scans challenging. Here, we are primarily interested in the rate of vortex breakdown, rather than absolute vortex strength. As such, the signature of the tip vortices in the streamwise velocity spectrum is used as a measure of vortex breakdown. By considering only the streamwise velocity component, this method neglects the three-dimensional structure of the tip vortices. However, for the purpose of deriving scaling laws, we consider this approach to be sufficient.

\begin{figure*}
    \centering
    \includegraphics[width=\textwidth]{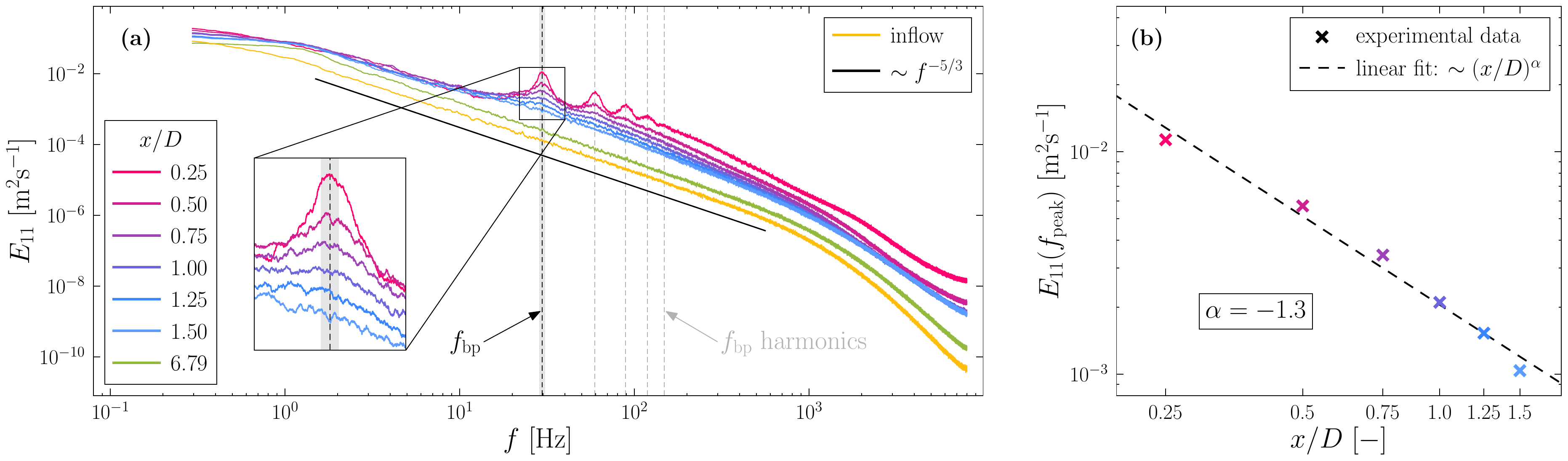}
    \caption{(a) inflow spectrum and wake spectra at the top of the rotor for the reference case at different downstream positions. $f_\mathrm{bp}$ is the blade pass frequency. The interval shown around $f_\mathrm{bp}$ corresponds to the region, in which the maximum value is taken for the determination of the peaks. (b) local maxima of the wake spectra around the blade pass frequency.}
    \label{fig:wake_spectra}
\end{figure*}

\autoref{fig:wake_spectra}a shows the velocity spectrum of the inflow at the rotor position as well as in the wake at different downstream positions for the reference case. In the wake, the $-5/3$ scaling in the inertial range is disrupted by the signature of the tip vortices. The highest deviation occurs close to the rotor, where the tip vortices cause the spectrum to exhibit a distinct peak at approximately the blade pass frequency~$f_\mathrm{bp} = 3f_\mathrm{rot}$, where $f_\mathrm{rot}$ is the rotational frequency of the wind turbine. This is the frequency at which tip vortices are created at a specific azimuthal position along the rotor edge and thus the frequency at which they pass by a specific measurement location in the wake. Close to the rotor ($x/D \lesssim 0.5$), peaks also occur at frequencies equal to an integer multiple of the blade pass frequency. These peaks correspond to the harmonics of the first peak. The height of the first peak decreases with distance from the rotor. These signatures of the tip vortices in the wake velocity spectra have previously been observed in the literature~\cite{pique_laboratory_2022, pique_dominant_2022, neunaber_distinct_2020,biswas_effect_2024, zhang_near-wake_2012}.

We consider the height of the peak at~$f_\mathrm{bp}$ to be related to the strength of the tip vortices (see~\autoref{sec:results:reference_case}). Thus, the maximum of the smoothed velocity spectrum within a~\SI{2}{\hertz} interval around the blade pass frequency is taken as a measure for the strength of the vortices. The width of this interval is chosen such that the slope of the spectra does not significantly affect the determined peak height even for small peaks. Details regarding the sensitivity of the peak height to the computation of the spectra are discussed in~\aref{sec:appendix_parameters}. The resulting peak heights as a function of downstream position for the reference case are shown in~\autoref{fig:wake_spectra}b and further elaborated in ~\autoref{sec:results:reference_case}.

\section{\label{sec:results}Results and Discussion}
\subsection{\label{sec:results:spectra_evolution}Evolution of the wake spectra}
The streamwise velocity spectra for the reference case are shown in~\autoref{fig:wake_spectra}a. They contain the distribution of turbulent kinetic energy in the streamwise velocity component~$E_{11}$ across the frequencies~$f$ that are present in the flow. The frequencies can be associated with the scales of the turbulence: low frequencies correspond to large scales, while high frequencies correspond to small scales. The inflow spectrum was obtained at the streamwise position of the rotor after removing the turbine, while the wake spectra were measured at $0.25 \leq x/D \leq 6.79$ downstream of the rotor. The inflow spectrum exhibits an inertial range of almost three decades, as shown by the $-5/3$ scaling, which is characteristic of high Reynolds number turbulence.

The downstream evolution of the wake spectra can be divided into two regions. 
In the first region $(x/D \lesssim 1.50)$, the wake spectra are dominated by the presence of the tip vortices, which appear to primarily affect the turbulence at scales equal to and smaller than the vortices themselves. As such, the spectral energy in the range~$\SI{e1}{\hertz}\lesssim f \lesssim\SI{e3}{\hertz}$ is higher than in the inflow, but decreases with downstream position.
At frequencies lower than the blade pass frequency (larger scales), the spectral energy is also higher in the wake than in the inflow, which might be due to generation of turbulent kinetic energy by the shear layer surrounding the wake. Interestingly, over~$\SI[parse-numbers=false]{10^0}{\hertz}\lesssim f \lesssim\SI{e1}{\hertz}$, which is within the inertial range of the inflow spectrum but lower than the blade pass frequency, the spectral energy remains constant with downstream position. Only the lowest frequencies ($f \lesssim \SI[parse-numbers=false]{10^0}{\hertz}$) show a decrease of spectral energy with downstream position, indicating a decay of these very large scales. This suggests that the turbulence at scales larger than the tip vortices is in equilibrium and slowly decaying, while the smaller scales are out of equilibrium due to the turbulence generated by the vortex breakdown.
At the highest frequencies~($f\gtrsim \SI{e3}{\hertz}$), which extend into the dissipation range, the measurements closest to the rotor ($x/D \leq 0.5$) show the highest spectral energy, but interestingly the spectra of the more downstream positions collapse. However, due to the use of $\mathrm{SF}_6$ to achieve Kolmogorov length scales $\sim \SI{e-5}{\metre}$, the dissipation range was not fully resolved in these measurements.

In the second region $(x/D \gtrsim 1.50)$, the signature of the tip vortices is no longer visible in the spectra, which are now similar in shape to the inflow spectrum. Presumably, the tip vortices have entirely broken down. With increasing distance from the rotor, the spectra are shifted downwards indicating an overall decay of turbulent kinetic energy. Far from the rotor~($x/D=6.79$) the spectrum resembles the inflow spectrum over the entire inertial and dissipation ranges, but is still shifted upwards, indicating that the wake turbulence has not fully decayed. This is in agreement with previous studies of wake turbulence, e.g.~\cite{gambuzza_influence_2023}. In the remainder of this manuscript, we focus on the first region, in which the tip vortices are still visible in the spectra.

\subsection{\label{sec:results:reference_case}Tip vortex breakdown}
The inset in~\autoref{fig:wake_spectra}a shows a close-up of the peaks around the blade pass frequency in the streamwise velocity spectra. In the absence of flow visualization, it is not possible to directly relate these peaks to specific vortex behavior or breakdown mechanisms. Nevertheless, their gradual decay may serve as a relative measure of the rate of vortex breakdown in various flow conditions. The frequency at which these peaks are located in the spectrum is related to the rate at which the vortices appear at the measurement location, rather than their size. At $x/D = 0$, a tip vortex is generated every time a blade passes by the reference location, so that we expect the vortices to appear in the velocity signal as a periodic disturbance with a frequency of exactly $f_{bp}$. As they are convected downstream and interact with the surrounding flow, we expect them to wobble and gradually break up so that the rate at which they appear at a given measurement location begins to fluctuate. This would cause the peaks to widen with increasing downstream position, which is what we observe in~\autoref{fig:wake_spectra}. As such, we interpret the magnitude of the peak to be a measure of the coherence of the vortex and the regularity of its appearance at the measurement location: a flatter and wider peak implies a less coherent vortex or larger fluctuations in its location, both of which can be associated with vortex breakdown.

To investigate the rate of tip vortex breakdown, we quantify the height of the aforementioned peak in the spectrum at each downstream position as described in~\autoref{sec:methodology:quantifying_tip_vortex_strength}. \autoref{fig:wake_spectra}b shows the downstream evolution of the peak height for the reference case. It appears that the peak height may follow a linear trend in this logarithmic plot, which would imply a power law scaling. In a previous study of wind turbine tip vortex breakdown, Hu et al. similarly observed a power law scaling of the vorticity of the vortex, quantified using particle image velocimetry~\cite{hu_dynamic_2012}. Such a power law scaling indicates that within the wake region studied here, the underlying process of tip vortex breakdown is self-similar and governed by a consistent physical mechanism. Further study is necessary to definitively confirm the linear nature of this relationship. Nevertheless, we here follow Hu et al. in the assumption that within the wake region studied here the breakdown process can in fact be described by a power law scaling of the form ~$E_{11}^\mathrm{peak}(\xi)= c\,\xi^\alpha$ where $E_{11}^\mathrm{peak}=E_{11}(f_\mathrm{peak})$ and $\xi=x/D$. The parameter~$c$ is the peak height at~$\xi=1$ and thus describes the vertical offset of the data. The parameter~$\alpha$ is the slope of the linear fit in the logarithmically-scaled graph and thus describes the rate of peak decrease. Since this scaling describes a decay process, $\alpha$ is negative and $E_{11}^\mathrm{peak}$ is undefined at $x=0$. Thus, for very small $x/D$ the above scaling cannot hold. The parameter $c$ may contain information about this initial stage, though we cannot determine from our measurements where exactly the power scaling takes over. In the following we investigate how variations in the flow conditions affect these scaling parameters, which are given in~\autoref{tab:alpha}.

%

\subsection{\label{sec:results:shear_ti}Effects of shear and turbulence intensity}
\begin{figure}
    \centering
    \includegraphics[width=\columnwidth]{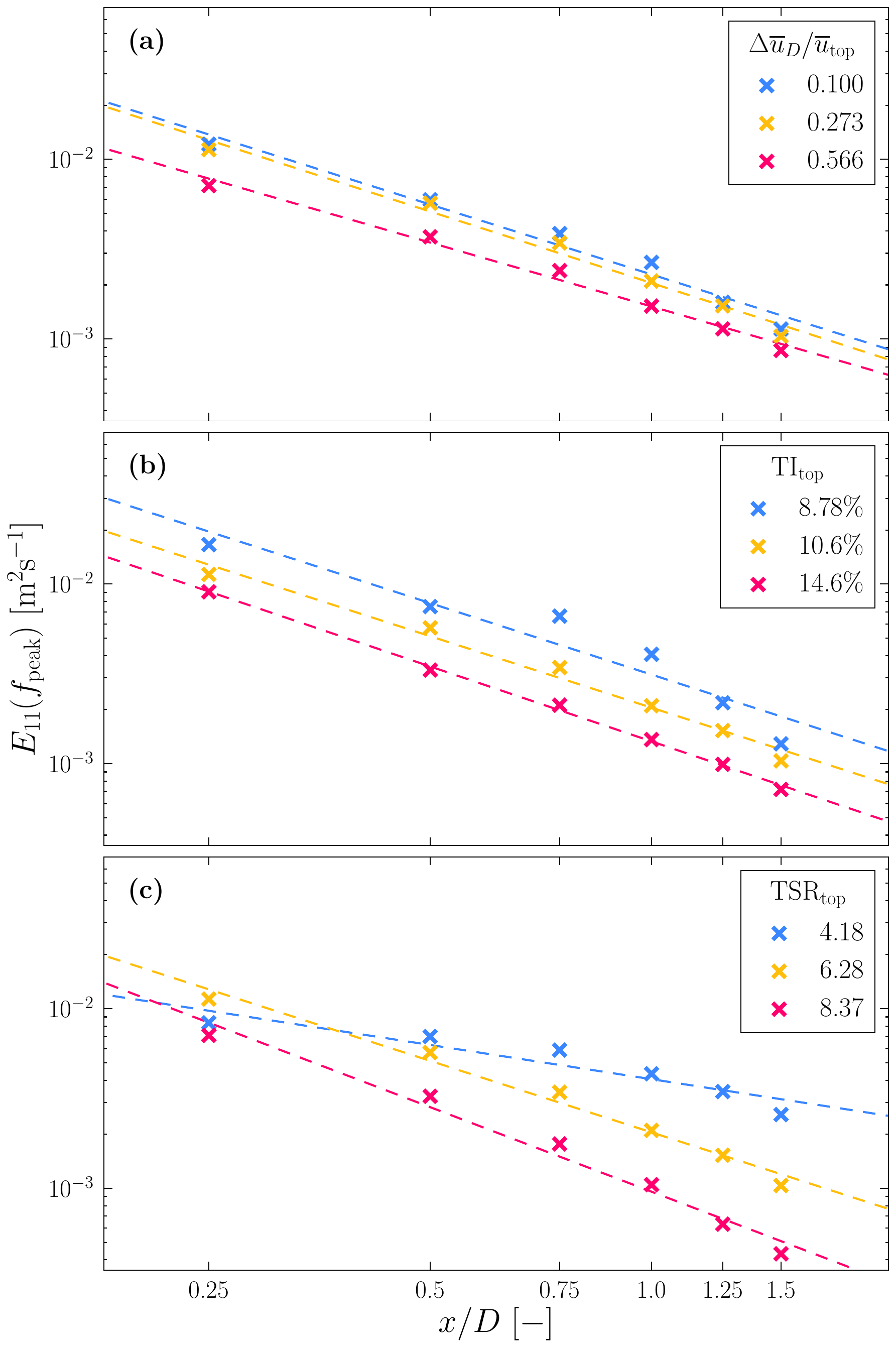}
    \caption{Spectral peak heights corresponding to tip vortices at different downstream positions for varying~(a) shear, (b) turbulence intensity and (c) tip speed ratio. The dashed lines correspond to power law fits.}
    \label{fig:peak_values_no_correction}
\end{figure}

\begin{table}
    \centering
    \begin{ruledtabular}
        \begin{tabular}{lcc}
             ID & $\alpha$ & $c$ $[10^{-3}\si{\metre\squared\per\second}]$ \\
            \hline
            low shear                   & \SI[multi-part-units=repeat]{-1.3(2)}{}   & \SI[multi-part-units=repeat]{2.3(2)}{}\\
            low turbulence intensity    & \SI[multi-part-units=repeat]{-1.3(3)}{}   & \SI[multi-part-units=repeat]{3.2(5)}{}\\
            low tip speed ratio         & \SI[multi-part-units=repeat]{-0.6(2)}{}   & \SI[multi-part-units=repeat]{4.1(4)}{}\\
            reference case              & \SI[multi-part-units=repeat]{-1.32(9)}{}  & \SI[multi-part-units=repeat]{2.0(2)}{}\\
            high tip speed ratio        & \SI[multi-part-units=repeat]{-1.6(2)}{}   & \SI[multi-part-units=repeat]{0.96(8)}{}\\
            high turbulence intensity   & \SI[multi-part-units=repeat]{-1.39(4)}{}  & \SI[multi-part-units=repeat]{1.33(4)}{}\\
            high shear                  & \SI[multi-part-units=repeat]{-1.18(7)}{}  & \SI[multi-part-units=repeat]{1.52(7)}{}\\
        \end{tabular}
    \end{ruledtabular}
    \caption{Parameters of the power law fit $E_{11}^\mathrm{peak}=c\,\xi^\alpha$ for the downstream evolution of the peak heights in the wake spectra.}
    \label{tab:alpha}
\end{table}
Variations in mean velocity shear (see~\autoref{fig:peak_values_no_correction}a) and turbulence intensity (see~\autoref{fig:peak_values_no_correction}b) primarily lead to a vertical shift of the fitted curves, reflected in the parameter $c$ in~\autoref{tab:alpha}. The vortex breakdown rate $\alpha$ remains relatively constant. As such, the peak heights are already different $x/D=0.25$ downstream of the rotor. This behavior is most striking for the variation in turbulence intensities and may indicate that within the wake region studied here, the underlying process of the tip vortex breakdown is only weakly affected by the inflow shear and turbulence intensity. Instead, their impact on tip vortex breakdown might occur very close to the rotor, during or immediately after vortex formation, thereby affecting the overall time it takes the vortices to break down. Interestingly, there is no notable difference in the breakdown process between the low and medium shear conditions, while the high shear condition seems to accelerate the breakdown process very close to the rotor. Because mean velocity shear is a relatively large-scale flow feature compared to the scales of the tip vortices, they may only feel the shear at short time-scales if it is sufficiently strong locally.

\begin{figure}
    \centering
    \includegraphics[width=\columnwidth]{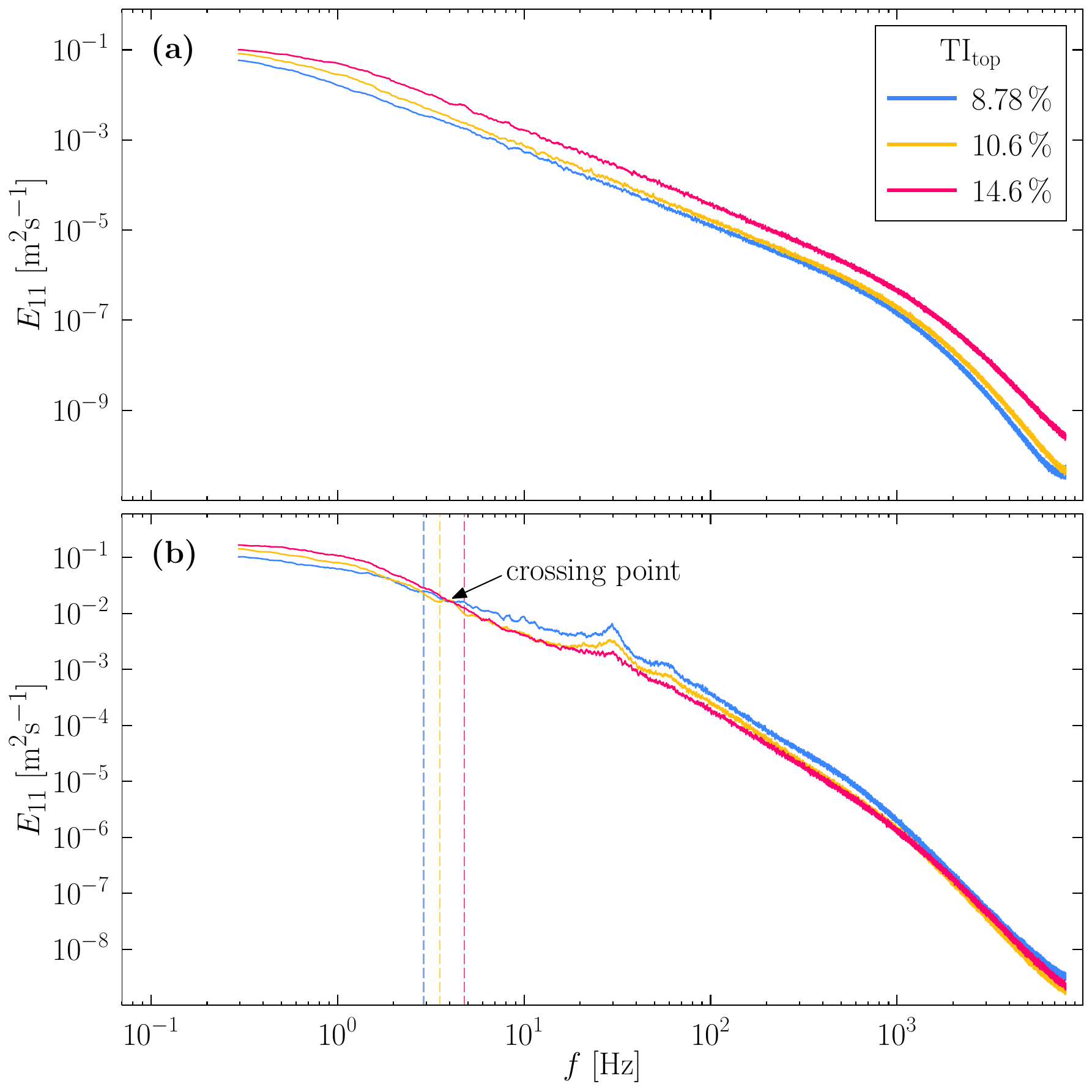}
    \caption{(a) Inflow spectra and (b) wake spectra at $x/D=0.75$ for different turbulence intensities in the inflow. The dashed lines correspond to the integral scales.}
    \label{fig:turbulence_intensity_crossing_point}
\end{figure}

The scales at which the turbulence interacts with the tip vortices is best seen in the velocity spectra. \autoref{fig:turbulence_intensity_crossing_point} shows inflow and wake spectra for the conditions with varying turbulence intensity. As expected, the inflow spectra are shifted vertically: higher turbulence intensity leads to more spectral energy at all scales, but interestingly it also reduces the integral scale (see~\autoref{tab:inflow}). Nevertheless, the integral scale is always on the order of the rotor diameter. 

At the low frequencies (large scales), the wake spectra follow the same trend as the inflow spectra, though the spectral energy is higher for all cases, presumably due to turbulence generated by the shear layer. However, at scales close to the integral scales, a crossing point occurs after which the trend is reversed, so that at the scales of the tip vortices, higher turbulence intensity leads to less spectral energy. In the dissipation range, the spectra collapse. This may indicate that it is primarily the large scales of the turbulence that interact with the vortices to facilitate their breakdown and the decay of the turbulence generated by this breakdown. The observed crossing point and the spectral energy at the large scales remain relatively constant throughout the range observed here (up to $x/D = 1.50$), which raises the question whether the turbulence generated by the vortex breakdown inhibits the decay of the large-scale turbulence generated by the shear layer. This shows the importance of considering which scales are present in the turbulence and how they interact with the tip vortices, rather than using only the turbulence intensity to characterize the flow.

\subsection{\label{sec:results:tsr}Effects of tip speed ratio}
The tip speed ratio appears to have a notable effect on the tip vortex breakdown rate~$\alpha$: higher tip speed ratios lead to faster breakdown, as shown in~\autoref{fig:peak_values_no_correction}c. Interestingly, at $x/D=0.25$ the peak heights are relatively similar for all three tip speed ratios investigated. This is particularly surprising because lower tip speed ratios are expected to induce slightly stronger initial tip vortices~\cite{hu_dynamic_2012}. As described in the introduction, lower tip speed ratios also result in larger spacing between subsequent vortices, which may lead to slower breakdown. This is consistent with the present data. 
\section{\label{sec:conclusion}Conclusion}

We conducted the first laboratory study of a wind turbine wake in a high Reynolds number~($\mathrm{Re}_D\sim 10^6$) turbulent flow. We demonstrate that the combination of a pressurized wind tunnel and an active turbulence grid allows for the recreation of realistic atmospheric boundary layer flow characteristics in a controllable and repeatable way. This enables us to investigate specific flow parameters like mean shear and turbulence intensity in isolation, while maintaining dynamic similarity with full-scale wind turbine flows.

We investigated the effect of various flow conditions on the scaling of the tip vortex breakdown process. The inflow conditions considered here were significantly more turbulent than those of most previous investigations of this phenomenon, so that interactions of the turbulence with the tip vortices were expected to be important for the breakdown process. We show that within the wake region investigated here, the breakdown process is described reasonably well by a power law scaling. The tip speed ratio has a significant effect on the breakdown rate throughout the investigated region, while turbulence intensity appears to primarily affect the breakdown process further upstream without affecting the breakdown rate in the observed region. The impact of mean velocity shear was overall less pronounced and did not notably affect the breakdown rate in the observed region. This points to the importance of the tip speed ratio for the tip vortex breakdown process even in a highly turbulent environment. The chosen tip speed ratio for wind turbines is commonly based solely on considerations regarding the power output of the singular turbine. For wind farms, in which re-energization of the turbine wakes is crucial, the aggregated power output might benefit from choosing slightly larger tip speed ratios, as these would lead to a faster tip speed breakdown.

We further show that the different length scales in the turbulence interact differently with the tip vortices and the largest scales are possibly most important for tip vortex breakdown. The turbulence intensity as a single parameter is unlikely to accurately capture this complexity. Rather the relationship between integral scales and vortex scales should be investigated further. Future work should also determine over what range, if at all, the power law scaling applied here is valid.

Our results underline the importance of the inflow conditions for the tip vortex breakdown process. Even small changes in turbulence intensity impact how far downstream the signature of the tip vortices persists in the wake, so that scalings observed under laminar inflow conditions are unlikely to accurately describe the vortex breakdown process in highly turbulent atmospheric conditions. Under highly turbulent conditions, tip vortex breakdown appears to be initiated very close to the rotor. Further study is necessary to investigate the effect of the inflow on the initiation of the breakdown close to the rotor.
\begin{table*}
    \begin{ruledtabular}
\begin{tabular}{lcrc}
ID & blockage & \multicolumn{1}{c}{$D_\mathrm{ou}$} & fan speed\\
\hline
\caseshearlow & [0.26, 0.26, 0.26, 0.26, 0.26, 0.26, 0.26, 0.26, 0.26] & 45000 & \SI{57}{\percent} \\
\casetilow & [0.00, 0.14, 0.32, 0.48, 0.55, 0.62, 0.69, 0.75, 0.81] & 3000 & \SI{52}{\percent} \\
\casebase & [0.00, 0.13, 0.31, 0.47, 0.54, 0.60, 0.66, 0.73, 0.79] & 9000 & \SI{51}{\percent} \\
\casetihigh & [0.00, 0.10, 0.30, 0.48, 0.64, 0.72, 0.79, 0.85, 0.91] & 50000 & \SI{51}{\percent} \\
\caseshearhigh & [0.33, 0.48, 0.62, 0.71, 0.79, 0.87, 0.93, 0.98, 1.00] & 2500 & \SI{49}{\percent}
\end{tabular}
    \end{ruledtabular}
    \caption{Parameter values for the active grid protocols. The blockage is given for each row of the active grid, from top to bottom. A fan speed of 100\% corresponds to \SI{1440}{\rpm}}
    \label{tab:active_grid_protocol_parameter_values}
\end{table*}
\section*{Data availability}
The data from this study are publicly available at DOI: \url{https://doi.org/10.17617/3.1EHZS9}

\begin{acknowledgments}
The authors would like to thank Julian Jüchter at the University of Oldenburg for adapting the model wind turbine to the experimental conditions of this study and for providing technical support throughout the experiments. We thank Michael Hölling and ForWind for providing us with the model wind turbine for this study. We thank Constantin Schettler for his assistance in designing the experimental setup and Andreas Renner, Andreas Kopp, Marcel Meyer and Artur Kubitzek for their support in setting up the experiment. We thank Eberhard Bodenschatz for helpful discussions during all stages of this study. 
\end{acknowledgments}

\appendix

\section{\label{sec:appendix_grid}Active grid protocols}
The optimization parameters for the grid protocols are the mean angle~$\overline{\phi}_i$ of each row $i$, the Ornstein-Uhlenbeck parameter~$D_\mathrm{ou}$ and the speed of the wind tunnel fan. The values of these parameters for each inflow case are presented in~\autoref{tab:active_grid_protocol_parameter_values}.
Here, instead of the mean angles~$\overline{\phi}_i$, the blockage~$B$ defined by $B_i=|\sin{\overline{\phi}_i}|$ is shown. A blockage of~0 corresponds to a fully opened row, a blockage of~1 to a fully closed one. The parameter~$\gamma_\mathrm{ou}$ of the Ornstein-Uhlenbeck process was set to~$\gamma_\mathrm{ou}=2$ for all protocols.

\section{\label{sec:appendix_parameters}Computation of the velocity spectra}
\begin{figure}
    \includegraphics[width=\columnwidth]{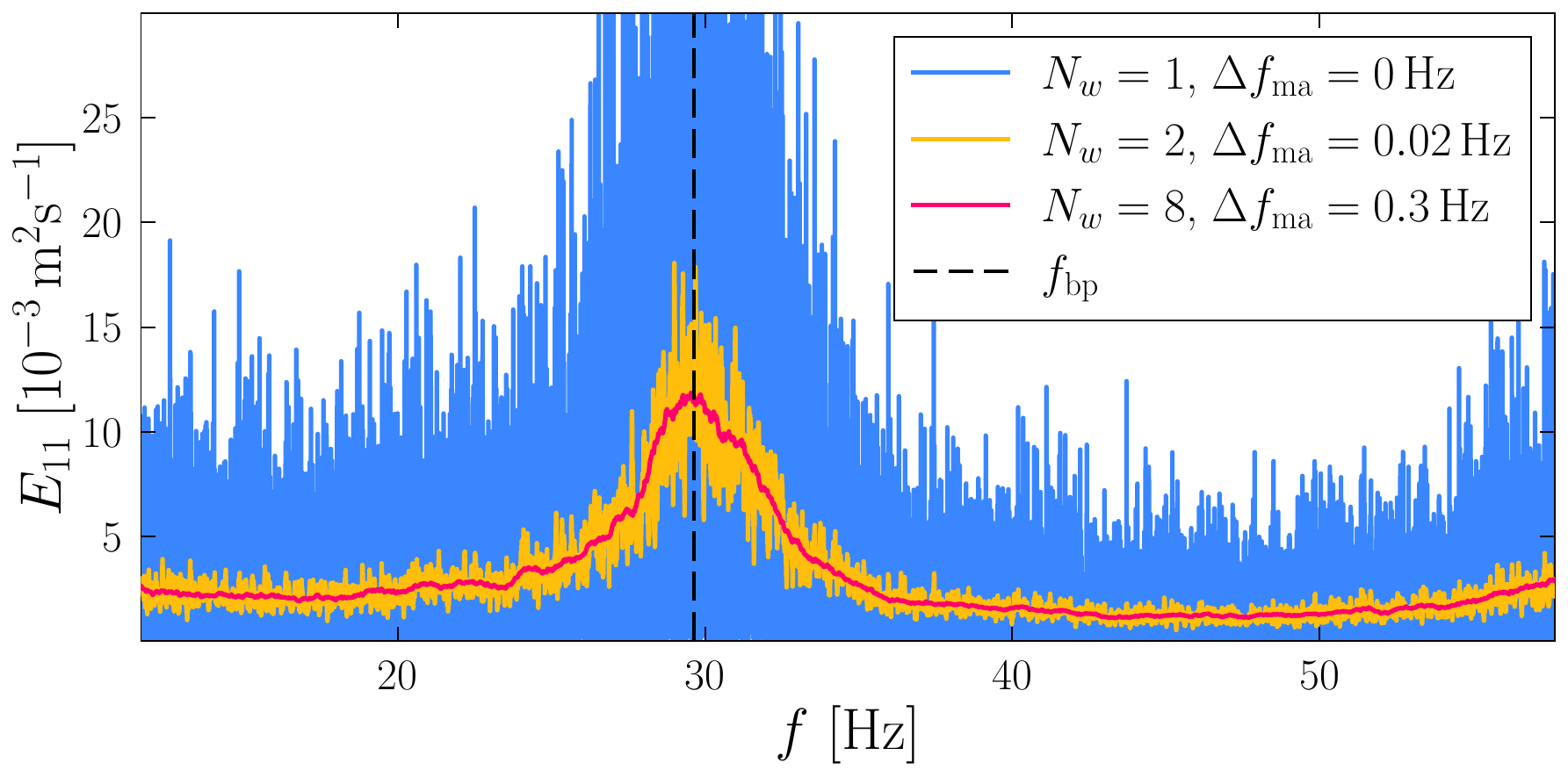}
    \caption{Spectrum computed with Welch's algorithm. On the raw spectrum (blue) no smoothing is applied. The other spectra are smoothed by using different combinations of the number of segments in Welch's algorithm and windowing width in a moving average filter.}
\label{fig:spectra_parameter_choices_comparison}
\end{figure}

The streamwise velocity spectra are computed using Welch's algorithm~\cite{welch_use_1967}. The algorithm divides the original time series into overlapping segments and computes the fast Fourier transform on each segment separately. The segments are chosen to have an overlap of~\SI{50}{\percent}. The number of segments is~$N_w$. The resulting spectra are smoothed using a moving average filter with a window width of~$\Delta f_\mathrm{ma}$. To demonstrate the effect of these parameters, \autoref{fig:spectra_parameter_choices_comparison} shows the raw spectrum for the reference case at~$x/D=0.25$ computed using Welch's algorithm with only one segment, as well as two other choices of $N_w$ and $\Delta f_\mathrm{ma}$.

\begin{figure}
    \includegraphics[width=\columnwidth]{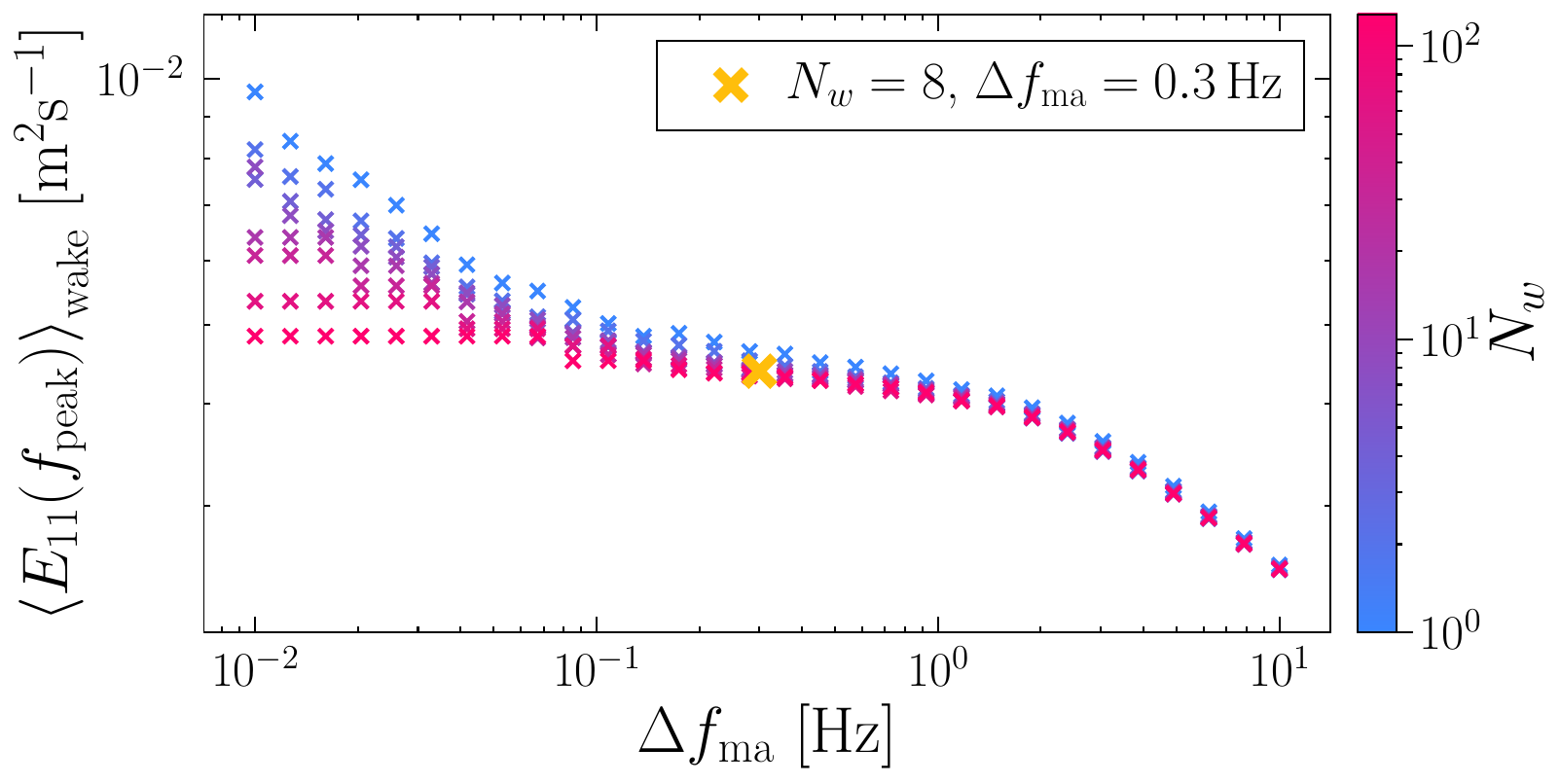}
    \caption{Streamwise-averaged peak height in the wake spectra around the blade pass frequency for different choices of the number of segments in Welch's algorithm~$N_w$ and the window width of a moving average filter~$\Delta f_\mathrm{ma}$. The yellow cross indicates the parameter combination used in this manuscript.}
    \label{fig:peak_heights_for_parameter_choices}
\end{figure}

Our analysis is based on the peak that arises at the blade pass frequency. Since we define the height of this peak to be the maximum value within a~\SI{2}{\hertz} interval around the blade pass frequency, the computed values differ significantly for the three parameter choices shown. Therefore, a parameter combination of $N_w$ and $\Delta f_\mathrm{ma}$ is desired, such that the evaluated peak height is relatively insensitive to the parameters used for its computation. \autoref{fig:peak_heights_for_parameter_choices} shows the peak height of the reference case averaged over the wake positions $0.25\leq x/D \leq 1.5$ as a function of the two smoothing parameters.

At $10^{-1}\lesssim\Delta f_\mathrm{ma} \lesssim 10^{0}$ and $N_w \gtrsim 4$ the peak height is sufficiently independent of both parameters. Thus, this regime provides robust parameter combinations for the computation of the spectra. For the analysis presented in this manuscript, $N_w=8$ and $\Delta f_\mathrm{ma}=\SI{0.3}{\hertz}$ were used.

\bibliography{references}

\end{document}